# Theoretical and experimental study of a new single-coil superconducting miniundulator

J. Kulesza[1], D. Waterman[1], N. Roscup[1], C.Z. Diao[2], A. Deyhim[1], H.O. Moser[2,3]

[1]*Advanced Design Consulting USA, Inc., 126 Ridge Road, Lansing, NY 14882, USA*

[2]*Singapore Synchrotron Light Source (SSLS), National University of Singapore (NUS), 5 Research Link, Singapore 117603*

[3]*Karlsruhe Institute of Technology (KIT), Network of Excellent Retired Scientists (NES) and Institute of Microstructure Technology (IMT), Postfach 3640, D-76021 Karlsruhe, Germany*

**Abstract**

The first pre-prototype of a single-coil superconducting miniundulator has been built and studied. Its basic specifications include 10 main periods and two end compensation periods, a period length of 7 mm, and a gap of 2 mm. The design is based on a racetrack-like coil configuration that is subsequently compressed to form the gap region with the spatially alternating currents flowing perpendicularly to the electron beam above and below the midplane. Operation up to an excitation current slightly beyond 400 A before quenching resulted in a peak magnetic flux density on axis of about 1 T and an undulator parameter of $K \cong 0.65$.

**Introduction**

Superconducting miniundulators (superminis) are being developed to achieve shorter period lengths and higher magnetic fields than possible with the prior state-of-the-art [1-8]. For a given electron energy of an existing or planned accelerator, superminis allow the generation of harder photons which is of interest for the upgrade or retrofitting of existing facilities. For a given photon spectrum, they enable significant reductions of electron energy and, thus, savings of accelerator cost with the additional benefit that the photon beam will be more brilliant in a storage ring because the emittance of an electron beam becomes smaller at lower electron energy. Superminis feature a volume in which superconducting wires are arranged transversely, e.g., in parallel perpendicularly, to the electron beam, separated by a gap, and excited by the same current that is spatially alternating from one wire package to the next one.

State-of-the-art is the double-coil supermini as first presented in 1998 by Hezel et al. [3]. Following this experimental demonstration, the concept was developed to finally achieve routine operation of a double-coil supermini in the 2.5 GeV ANKA storage ring at KIT [4-6]. Additional work has been done along these lines at ANL [7] and LBNL [8] focusing on double-coil undulators.

Here, we present a radical simplification of the design of a supermini that needs only one single coil [9,10]. Having less weight, size, and stored energy, it can be more easily integrated into accelerator structures. Engineering design, simulation, manufacturing, and first experimental results are shown that were achieved with a pre-prototype in a liquid helium bath cryostat. The pre-prototype has 10 normal and two modified end periods. The period length is 7 mm which would result in a fundamental photon energy of 8.5 keV produced by a 2.5 GeV electron beam. As this work was a proof-of-concept manufacturing study limited in budget and time, results successfully demonstrate the principle of the single-coil supermini without being optimized.

**Design and manufacturing**

The undulator is designed to have a period length of 7 mm, a gap of 2 mm and 10 periods plus one matching period at either end of the undulator (Fig. 1). Such parameters were chosen for exemplifying the technology, without having any specific accelerator application in mind. A rectangular cross section NbTi wire of 0.726×1.206 mm$^2$ (Supercon SC-54S43) with 54 filaments of NbTi of a nominal diameter of 94 μm is used.

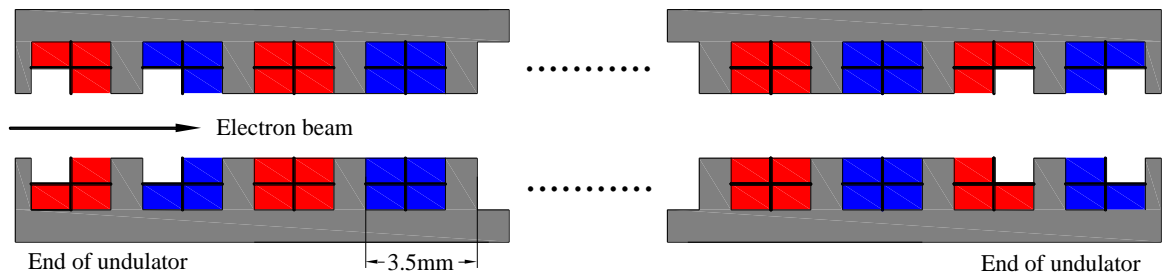

Fig. 1: Schematic of wire packages as built in main periods and end matching periods.

In a given wire package, the wire is wound 2 times to form the bottom layer and a further 2 times for the top layer such that the whole cross section comprises 2×2 wires. For a 10-period undulator, there are 20 4-wire coils plus 4 matching 3-wire coils at both ends of the undulator resulting in 25 poles of the ferromagnetic pole piece. For ease of manufacturing, the matching coils were simplified compared to the design [10] by featuring two wires at the groove bottom and one wire on top as represented in Fig. 1.

The coil is first wound on two non-ferromagnetic bobbins in a solenoidal shape (Fig. 2 a)) and then compressed to a dog bone shape by inserting two ferromagnetic pole pieces to form the field region of the gap. The specification includes 50 mm diameter of bobbins, 100 mm and 50 mm of width and height of the pole pieces, respectively, a length of the wired region on bobbins and pole pieces of 84.96 mm, and a gap width of 2 mm. As illustrated in Fig. 2 a), a single superconducting wire is wound under controlled mechanical tension around two spools or bobbins in a racetrack geometry. Consisting of 54 filaments of NbTi embedded in copper, with a Formvar jacket, the wire is manufactured by SuperCon under part number SC-54S43. Note that while the tension on the SC wire is significant during winding, the wire can easily withstand high mechanical tension. It will continue to be superconducting until the superconducting fibers deform to a point where they are no longer continuous. This occurs only at extreme tension. In fact, rectangular SC wire is formed from circular wire demonstrating a high tolerance to strong mechanical deformation. None-the-less, to be on the safe side, the winding mechanism includes a tension monitor.

The bobbins are machined with partial grooves to accept the wires which are rectangular (0.726 mm × 1.206 mm). The wires are wound around both bobbins with 4 turns per groove – two end section grooves have only have 3 turns. Figure 3 shows an actual bobbin of the pre-prototype during construction. Bobbins, pole pieces, and holding elements are manufactured in the precision machine shop, partly involving CNC machining. Figure 4 shows photos of a test winding of Cu wire on the bobbins that are held by the winding frame. The grooves are 2.6 mm wide and have a pitch of 3.5 mm in the pre-prototype. The bobbins are wound in such a way that, within a single turn, a wire in the top groove returns on the adjacent groove on the bottom. The bobbins are entirely wound in the forward direction first and then the reverse. Four turns are made on each full groove before moving to the next "forward" groove skipping a "reverse" groove. The wires are parallel between the bobbins, but must crossover each other on the outside of the bobbin in order to fill alternating grooves. The wire is reversed at a turning nut in order to subsequently fill reverse grooves. To provide a separate and distinct groove that supports the wire all the way around the outside of the bobbin is a considerable task. Grooves must be cut in an "S" pattern on a curved surface at varying depths. However, in this pre-prototype, no attempt was made at producing a point-to-point (P-P) tool path that a CNC machine could execute. Instead, a smooth outer bobbin surface was provided. The forward wires were wound and then potted with epoxy before winding the reverse wires which were also potted. This proved to be a successful approach eliminating the complicated task of producing a mathematical model from which a tool path could be generated.

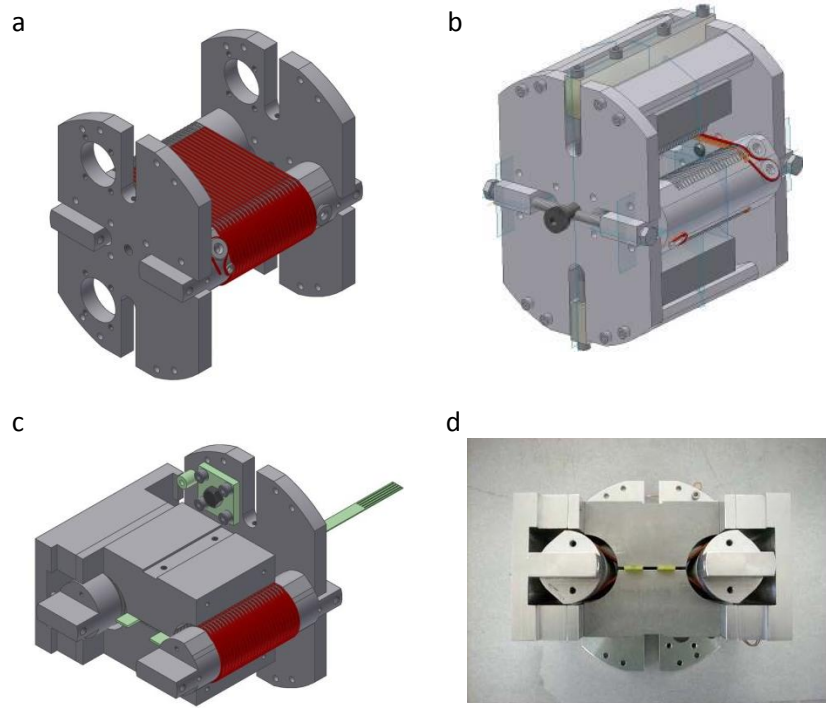

Fig. 2: Schematic of coil winding and compression. a) Superconducting wire wound on two bobbins held by holding plates. b) Assembly of bobbins with holding plates and pole pieces before compression shown without most of the wire for clarity. c) Drawing of the fully assembled supermini after mounting pole pieces and completing compression. d) Photo of supermini after completion of compression and clamping.

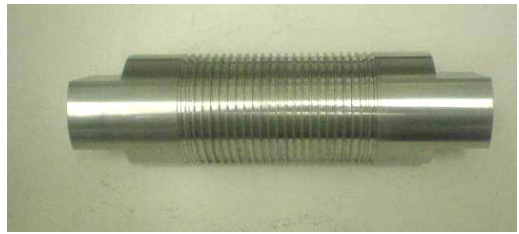

Fig. 3: Actual bobbin showing superconducting wire grooves. Diameter of a bobbin 50 mm, overall length 178 mm, wired region length 84.96 mm.

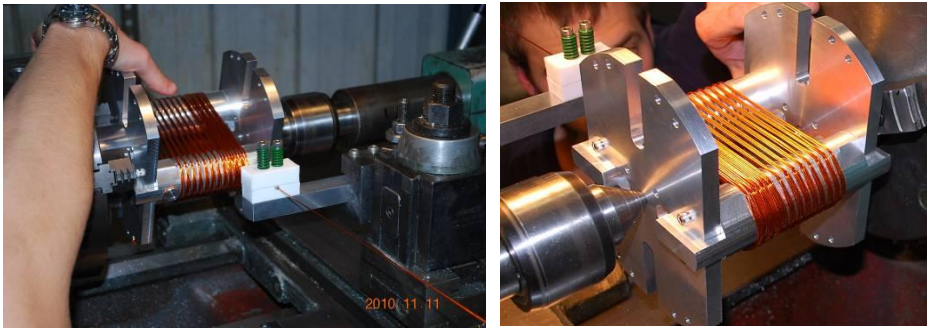

Fig. 4: Winding set up showing bobbins and holding frame mounted on a lathe for a Cu wire test winding.

Once the bobbins have been wound and the outside potted, pole pieces made of ferromagnetic material, in occurrence, ASTM 1018 steel, are mounted above and below the bobbins on the same winding frame (Fig. 5). Pole pieces also have grooves to accept the SC wires as shown in Fig. 5 a). These grooves are the ones schematically shown in Fig. 1. The bottom of the groove is slightly crowned in the center while the top of the ridge between grooves is flat and slightly chamfered. This is done to gently assist the wires into the grooves during transformation and to maintain their firm mechanical contact to the groove bottom. The cross-sectional contour of the pole pieces is shown in Fig. 5 b).

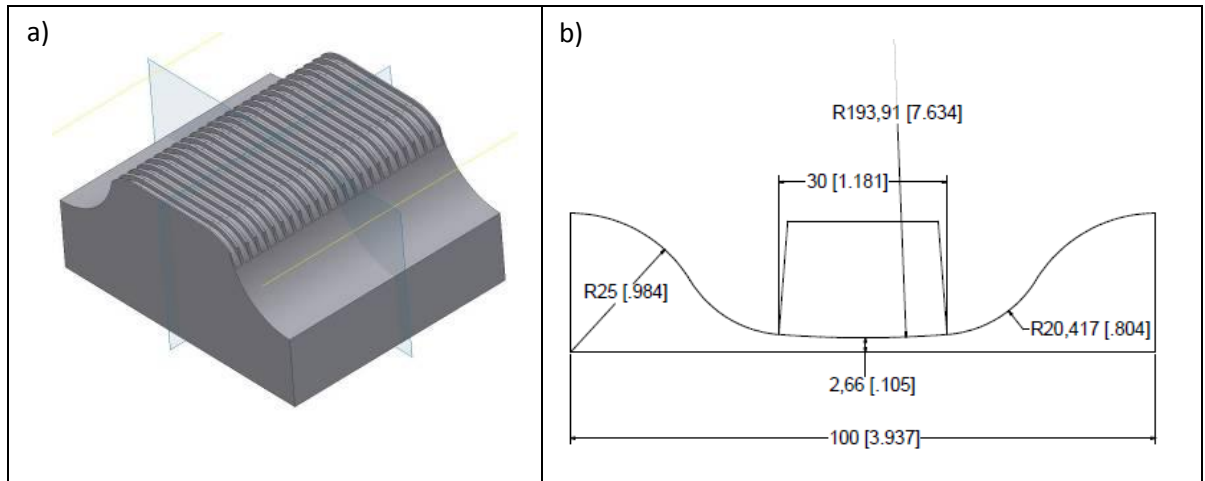

Figure 5: a) Pole piece. b) Cross-sectional contour of pole pieces that is identical with the wire path. Numbers in mm except for numbers in square brackets that are given in inches.

Figure 2 b) displays a drawing of the fully assembled pieces including the pole pieces before transformation. One can also see the direction reversing nuts at one end of the front spool. This winding process was done manually for this pre-prototype. Obviously, it would be automated for longer devices. From that status, pole pieces and bobbins are moved towards each other in a controlled way such that pole pieces are approaching to form the gap region and bobbins are approaching as well to enable the motion of pole pieces while maintaining the winding under mechanical tension (Fig. 2 c) and d)). The resulting pre-prototype of a 10-period supermini is shown in Fig. 6.

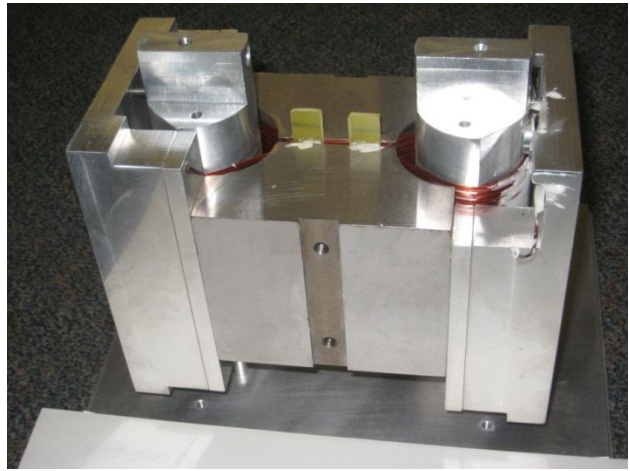

Fig. 6: Photo of final supermini after compression.

**Magnetic field calculations**

Field calculations were performed by means of RADIA [11] assuming parallel wires over a wide enough range such as to neglect the influence of wire return loops (Fig. 7). The iron used in the simulation was the RADIA pre-programmed material "RadMatSteel42" which is a low carbon steel with C < 0.19% having magnetic properties close to ASTM 1018. The magnetic flux density B is plotted versus the z-coordinate which describes the longitudinal direction of the supermini. The current in the coil was 400 A. It can be seen that, by design, the shape of the field is not perfect. The peak fields of the different poles are not equal and there is a certain positive dipolar field contribution in the undulating range. As shown in [12], these deformations are partly due to the finite length of the undulator and can be compensated by simple correction coils.

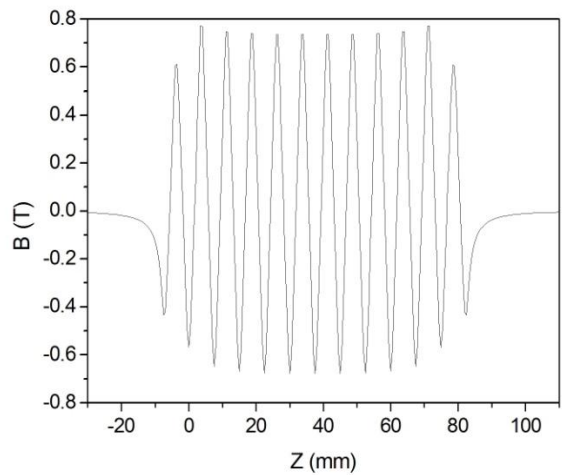

Fig. 7: Simulated vertical field component on axis at 400 A excitation current.

Figure 8 shows how the field increases from the midplane towards the poles. This illustrates that in order to generate a narrow line spectrum of radiation, the electron beam must be well centered on and confined to a central aperture of ± 200 µm around the midplane, a requirement in easy reach of modern

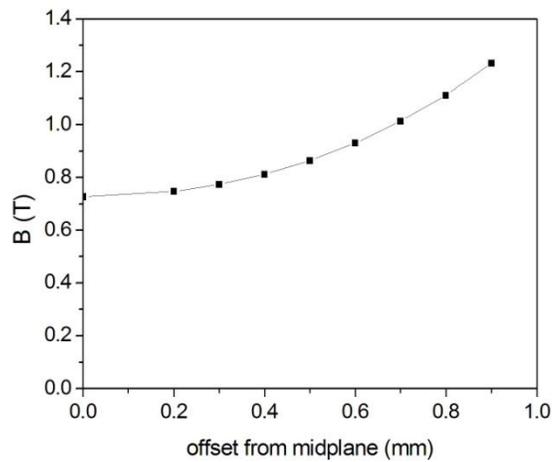

Fig. 8: Amplitude of vertical field component versus offset from the midplane.

3$^{rd}$ generation storage ring light sources. Figure 9 shows the vertical field component on axis versus excitation current illustrating the influence of the magnetization. At lower current values, the slope of B versus I is bigger due to the magnetization of the ferromagnetic pole piece. At around 400 A/mm$^2$, the magnetization reaches saturation and the further increase is only made by the H contribution of the coil. Such a behavior was demonstrated earlier by [13].

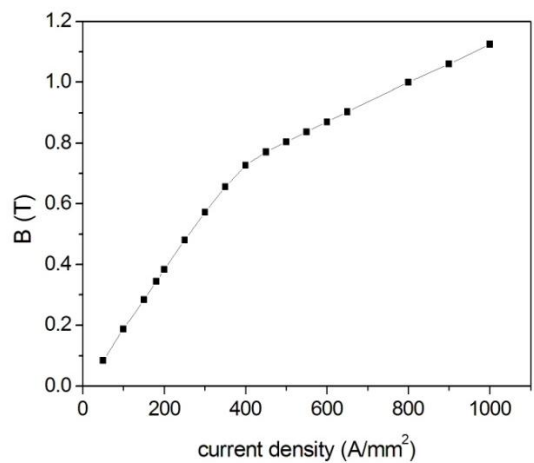

Fig. 9: Magnetic flux density versus exciting current density showing saturation of magnetization of the ferromagnetic pole piece at about 400 A/mm$^2$.

**Experimental set up and results**

Experiments were performed in a bath cryostat at liquid helium temperature. Dubbed Vertical Test Facility (VTF), it consists of a tall outer vertical framework that supports and guides an inner framework that holds the supermini. The latter is attached by the upper end plate to a central stainless steel tube that provides a path and a location for guidance features for the Hall probe which is moved vertically inside the supermini. Since the gap is only 2 mm, a set of fiberglass guides were mounted inside the supermini to guide the Hall probe which was mounted to a fiberglass wand. Very delicate Hall probe wires were epoxied to the wand, transitioned to heavier gauge Kapton coated wires, and carried up through a central fiberglass tube. Additional wires were installed for a thermal sensor mounted to the supermini. At the top, the outer stainless steel tube was connected to a long stroke vacuum bellow. The central fiberglass tube was mounted to the top of the bellow, and features inside the outer tube guided the central fiberglass tube. The wires were brought out top of the fiberglass tube, and the egress was sealed with epoxy. A stepper motor driven ball screw was used to move the central fiberglass tube compressing the bellows. An encoder mounted to the motor gave the position feedback. The connection for the power was provided by 16 mm diameter copper rods. The top (in-air) part of the copper rods were joined to large copper plates that were connected to four copper welding cables of 103 mm$^2$ cross section. The copper rods were encased in a 400 mm diameter stainless steel tube which was sealed between the top and cryostat lid and filled with liquid nitrogen to precool the copper rods. The rods extended down through the lid to the top of the supermini through a set of heat shield baffles. Copper fins mounted to the copper rods inside the cryostat expanded the surface area and made use of the evaporated helium in the upper part of the cryostat to pre-cool the copper rods. A cryo liquid level sensor was mounted down through the lid to the bottom of the supermini. The copper rods and central pipe were suspended from the cryostat lid. The cryostat was a Janis 9VSRD with a diameter of 230 mm and a depth of 1350 mm. A quench protection circuit was employed to shut down the power supply if a quench occurred. 60 liters of LHe provided about 1 hour of test time.

Magnetic flux density was measured by means of a low-temperature calibrated Hall probe (Cryomagnetics Cryogenic probe HSU-1) that could be moved along the undulator axis within the gap covering also some outside range on both ends. Figure 10 shows the completed supermini pre-prototype hanging in the VTF. Bus bars and the guide for the Hall probe are also visible.

The result of such a field measurement is given in Fig. 11. Raw data are shown for various values of the coil excitation current ranging from 50 A to 400 A. Although the quench current was expected to exceed 1 kA, the pre-prototype quenched consistently beyond 400 A indicating a systemic reason the cure for which is discussed below. In Fig. 11, the abscissa corresponds to the z axis with an increment of 0.1 mm from point to point. It can be seen that the field exhibits a number of 25 poles as 13 dips and 12 peaks with the polarities of poles as expected. Besides the characteristic oscillation of the undulator field, the data also indicate an upward shift of the magnetic field inside the undulator. This is a positive dipolar contribution of about 0.34 T confined to the extension of the pole pieces as can be seen from the zero baseline outside the undulator at either end of the graph. Its value can be found from the level at which the crossings of the magnetic field undulations are equidistant. This dipole field arises by design, namely from the cross-overs of the coils on the outside of the bobbins. A schematic winding map (Fig. 12) illustrates the longitudinal currents as arrows pointing to the right at the cross-overs of bobbin 1 and to the left at the cross-overs of bobbin 2. Assuming the magnetic properties of the pole pieces being described by RadMatSteel42 of RADIA and using the known geometry of the supermini, the dipole field shown in Fig. 13 is obtained. Obviously, the crossover current is the main cause of the dipole field. The maxima and minima of the undulating spatial field modulation in Fig. 13 represent the poles and the grooves in between, respectively. In order to achieve the pure undulator field, this dipolar field contribution can be compensated by a current loop running longitudinally down one bobbin and up the other one with the current flowing in opposite direction. Due to the limitations in time and budget of the present project, such a compensation loop was not installed. However, it will be part of subsequent prototypes.

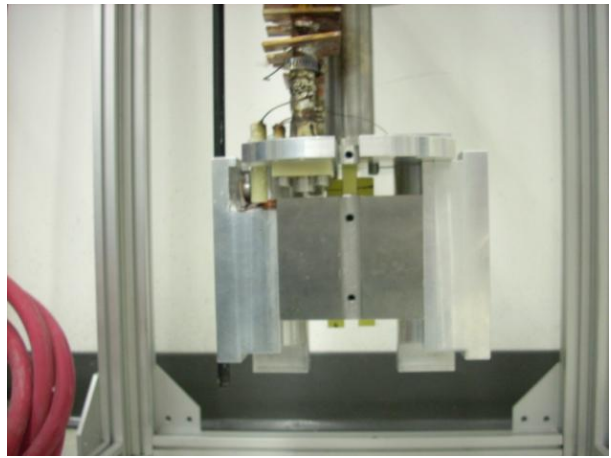

Fig. 10: Completed supermini to be suspended in liquid helium bath cryostat of the VTF. Movable Hall probe is used for measuring the longitudinal dependence of the magnetic field. Bus bars for the excitation current can be seen as well.

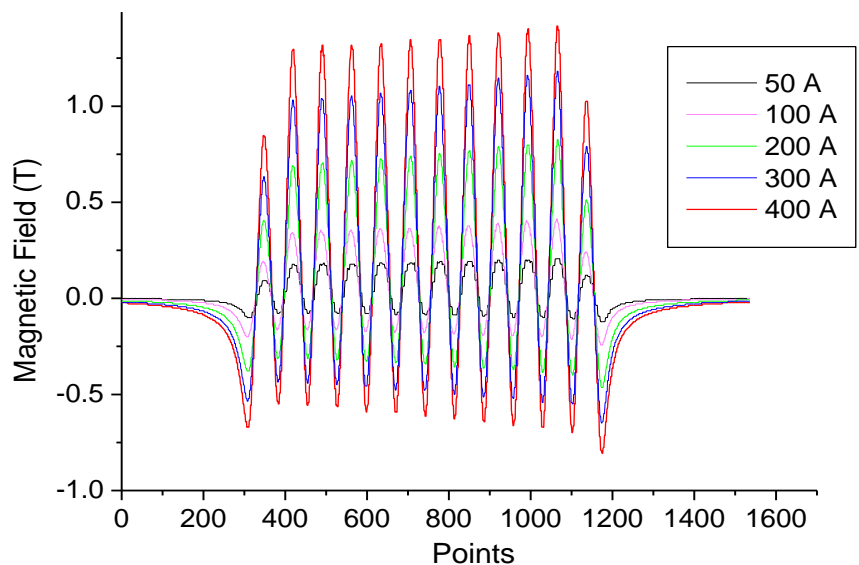

Fig. 11: Raw data of field measurement. Inset shows values of excitation current leading to different field amplitudes. Point-to-point increment is 0.1 mm.

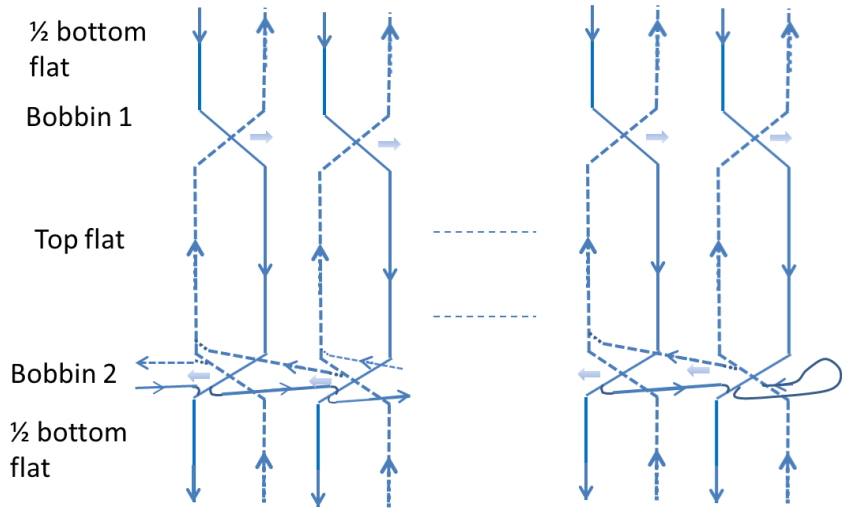

Fig. 12: Schematic winding map showing the origin of the longitudinal current (light solid arrows) from the wire crossings on the bobbins. This current generates the dipole field that shifts the undulator field up. It can be corrected by a wire loop with the opposite current.

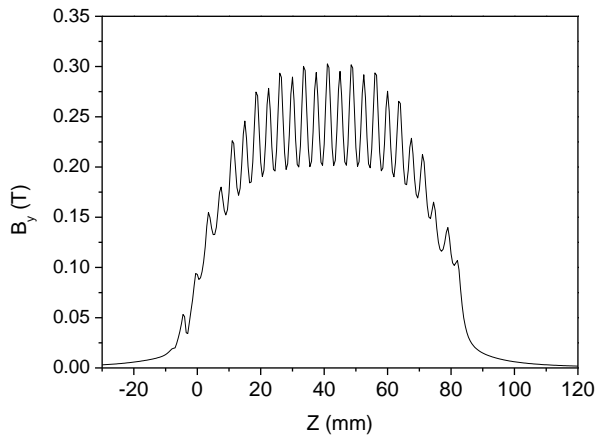

Fig. 13: Plot of dipole field created by crossover currents at 400 A excitation current.

Furthermore, the absolute value of the peak magnetic field is increasing from one end to the other due to the gap between pole pieces that is shrinking by an amount of 0.5 mm from one end to the other as built. Including these conditions, we present the simulated undulator field in comparison with the measured field in Fig. 14.

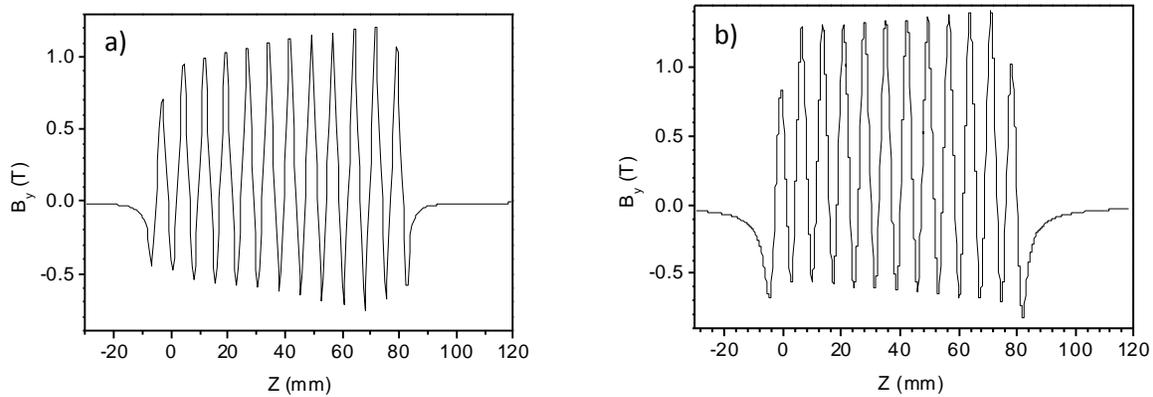

Fig. 14:  a) Simulation of the magnetic field including the crossover-current dipole field and the effect of the gap change along the undulator at 400 A excitation current. b) Measured raw field data as taken from Fig. 11.

**Discussion**

Correcting the raw field data for the unwanted dipole field caused by the coil crossings and taking into account the influence of the slight wedge shape of the gap lead to a reasonable agreement between measured and simulated field data. A further small contribution to the field shift in the undulator region also arises from the shape asymmetry of the undulator field owing to its finite-length without compensation or with incomplete compensation [12] as well as from asymmetries between forward and backward wiring between coils. These imperfections will be accounted for in the next prototype.

The quenches of the superconducting coil occurred at lower excitation currents than the wire would stand. It is believed that the causes include  that, for constructive reasons, there is a small section with a gap between bobbins and pole pieces in which the wires are not supported as can be seen from Fig. 2 d), and that the superconducting wires extending from coil to bus bars are not supported. This will also be improved in the next version. Upon successful implementation of such improvements in subsequent prototypes, the single-coil supermini is expected to outperform permanent magnet undulators at period lengths of about 7 mm or smaller.

A further effort needs to go into the magnetic field measurement system. The positioning of the Hall probe in y-direction perpendicular to the gap needs to be re-evaluated with the goal to know the y value rather accurately and to keep it constant during the scan, i.e., moving the Hall probe parallel to the midplane. The pitch and roll angles of the Hall probe also have to be minimized and controlled.

For the sake of completeness, we note that, besides considerations of winding and manufacturing, there are serious heat load and beam life time issues that the single-coil supermini shares with the double-coil species. Countermeasures applicable to both species would include carefully designed minimum beta sections, scrapers, and chicanes. In addition, they would take into account forthcoming results of detailed studies in progress at ANKA and collaborating teams [14,15].

**Conclusion**

In this proof-of-concept study, the principle of a single-coil supermini was theoretically and experimentally confirmed to be viable by designing, simulating, manufacturing, and operating a 10-period pre-prototype. A forthcoming prototype will benefit from improvements derived from the results of the present study. Single-coil superminis are smaller and less heavy, and promise to be less expensive than double-coil counterparts, thus facilitating their use in storage rings and linear accelerators. Depending on availability of superconducting wires, the period length may be further reduced to boost photon energy even higher.

**Acknowledgment**

This work was supported by the National Research Foundation of Singapore under Proof-of-Concept grant NRF2009NRF-POC001-119. Continuous support by M.B.H. Breese is gratefully acknowledged.

**References**

[1] I. Ben-Zvi, Z. Y. Jiang, G. Ingold and L. H. Yu, *Performance of a superconducting microundulator prototype*, Nucl. Instrum. Methods A **297,** 301(1990).
[2] H.O. Moser, B. Krevet and H. Holzapfel, *Mikroundulator*, Forschungszentrum Karlsruhe, German patent P 41 01 094.9-33, Jan. 16, 1991.
[3] T. Hezel, B. Krevet, H.O. Moser, J.A. Rossmanith, R. Rossmanith, and Th. Schneider, *A superconductive undulator with a period length of 3.8 mm*, J. Synchrotron Rad. **5,** 448-450(1998).
[4] T. Hezel, M. Homscheidt, H.O. Moser, R. Rossmanith, Th. Schneider, H. Backe, S. Dambach, F. Hagenbuck, K.-H. Kaiser, G. Kube, W. Lauth, Th. Walcher, *First beam test of a superconductive in-vacuo mini-undulator for future x-ray lasers and storage rings*, Free Electron Lasers 1999, J. Feldhaus, H. Weise, eds., pp. II-103, North Holland, 2000.
[5] S. Casalbuoni, M. Hagelstein, B. Kostka, and R.Rossmanith, *Generation of x-ray radiation in a storage ring by a superconductive cold-bore in-vacuum undulator,* Phys. Rev. ST Accel. Beam **9**, 010702 (2006).
[6] A. Bernhard, B. Kostka, R. Rossmanith, D. Dölling, A. Hobl, D.Krischel, S. Kubsky, U. Schindler, E. Steffens, *Planar and Planar Helical Superconductive Undulators for Storage Rings: State of the Art*, Proceedings of EPAC 2004, MOPKF025, pp. 354-356, Lucerne, Switzerland, 2004 (See http://accelconf.web.cern.ch/AccelConf/e04/PAPERS/MOPKF025.PDF).
[7] E.R. Moog, M. Abliz, K. Boerste, T. Buffington, D. Capatina, R.J. Dejus, C. Doose, Q. Hasse, Y. Ivanyushenkov, M. Jaski, M. Kasa, S.H. Kim, R. Kustom, E. Trakhtenberg, I. Vasserman, J.Z. Xu, N.A. Mezentsev, V.M. Syrovatin, *Development status of a superconducting undulator for the Advanced Photon Source (APS)*, Proceedings of IPAC 2010, WEPD047, pp. 3198-3200, Kyoto, Japan, 2010 (See http://accelconf.web.cern.ch/AccelConf/IPAC10/papers/wepd047.pdf).
[8] R. Schlueter, St. Marks, S. Prestemon, D. Dietderich, *Superconducting undulator research at LBNL*, Synchrotron Radiation News **17**(1), 33-37(2004).
[9] Herbert O. Moser, Caozheng Diao, *Single-coil superconducting miniundulator*, US Patent Application US 2011/0172104A1, Jul. 14, 2011
[10] C.Z. Diao, H.O. Moser, *Superconducting miniundulators*, in *Superconductivity and Superconducting Wires* (Horizons in World Physics, Volume 267), Dominic Matteri and Leone Futino, eds., pp. 321-332, Nova Science Publishers, Hauppauge, NY, USA, 2010.

[11] P. Elleaume, O. Chubar, J. Chavanne, *Computing 3D Magnetic Fields from Insertion Devices*, Proceedings of PAC 1997, pp. 3509-3511 (See http://epaper.kek.jp/pac97/papers/pdf/9P027.PDF).

[12] H.O. Moser, C.Z. Diao, *Finite-length field error and its compensation in superconducting miniundulators,* Nucl. Instr. and Meth. A **535,** 606-613(2004).

[13] H.O. Moser, R. Rossmanith, *Magnetic Field of Superconductive in-vacuo Undulators in Comparison with Permanent Magnet Undulators,* Nucl. Instr. and Meth. A **490,** 403-408(2002)

[14] S. Casalbuoni, A. Grau, M. Hagelstein, R. Rossmanith, F. Zimmermann, B. Kostka, E. Mashkina, E. Steffens, A. Bernhard, D. Wollmann, T. Baumbach, *Beam heat load and pressure rise in a cold vacuum chamber*, Phys. Rev. ST Accel. Beams **10,** 093202 (2007)

[15] M. Hagelstein, T. Baumbach, S. Casalbuoni, A. W. Grau, B. Kostka, R. Rossmanith, D. Saez de Jauregui, B. Diviacco, P. Elleaume, J. Chavanne, E. J. Wallén, E. Mashkina, A. Bernhard, D. Wollmann, *Design, development and testing of diagnostic systems for superconducting undulators*, Proceedings of PAC09, WE5RFP068, pp. 2417-2419, Vancouver, BC, Canada, 2009 (See http://accelconf.web.cern.ch/AccelConf/PAC2009/papers/we5rfp068.pdf).